# SCSPG (Semi-Circle Segmented Path Generator): How to Use and an Example in Calculating Work of Friction along Curved Path

S. Viridi[*] and S. N. Khotimah

Physics Department, Institut Teknologi Bandung, Ganesha 10, Bandung 40132, Indonesia
[*]dudung@fi.itb.ac.id

Abstract

A program called SCSPG (Semi-Circle Segmented Path Generator) is presented in this report. How it works is described and an example of it is illustrated using a case of work of friction along a curved path. As a benchmark for the program, work of friction along straight path is calculated and then compared to theoretical prediction.

Abstrak

Sebuah program bernama SCSPG (Semi-Circle Segmented Path Generator) dihadirkan dalam tulisan ini. Cara kerja program tersebut disajikan dan sebuah contoh penggunaannya diberikan dalam bentuk kasus perhitungan kerja gaya gesek sepanjang lintasan melengkung. Sebagai pembanding kinerja program, kerja gaya gesek sepanjang lintasan lurus dihitung dan kemudian dibandingkan hasilnya dengan prediksi teori.

**PACS**: 45.10.-b, 02.70.-c, 07.79.Sp , 07.05.Tp

## Introduction

Information of path details is needed in some problem such as mechanics [1] and fluids [2], where the application could be also fun [3]. Based on these needs an algorithm has been introduced [4], which is the core of SCSPG (Semi-Circle Segmented Path Generator) program. Work of friction a long a curved path for a point mass object under influenced of earth gravity is not a favorite problem in common physics textbooks. The major difficulty of this type of problem is how to define normal force that is always tangent to the curved path, since friction force is proportional to the normal force. Discretizing the curved path into semi-circle sub-paths could be the solution, since the normal force can be defined clearly in a semi-circle sub-path using concept of centripetal force.

## A curved path construction

### Semi-circle sub path

A curved path $s$ can be discretized into several sub-paths that have a semi-circle form, which is defined using six parameters. For a sub-path identified by index $k$, these parameters are initial position $x(0,k)$ and $y(0,k)$, length of sub-path $L_k$, initial angle $\theta(0,k)$ and final angle $\theta(N_k,k)$, and radius $R_k$, as shown in TABEL 1. Parameter $N_k$ is the number of unit length $\Delta s$ that constructs a sub-path $k$ through relation

$$N_k = \frac{L_k}{\Delta s}. \qquad (1)$$

Value of $\Delta s$ is chosen to be the same for all sub-paths in constructing the curved path $s$. An example of a semi-circle sub-path is given in FIGURE 1. The direction of the semi-circle sub-path is given by the arrow, which is counter-clockwise in this case. Direction of a sub-path is always from $\theta(0,k)$ to $\theta(N_k,k)$.

**TABLE 1**. Parameters for a sub-path.

| Parameters | Meaning |
|---|---|
| $x(0,k)$ | Initial position in $x$ direction for sub-path $k$ |
| $x(0,k)$ | Initial position in $y$ direction for sub-path $k$ |
| $L_k$ | Length of the sub-path $k$ |
| $\theta(0,k)$ | Initial angle measured from $x$ axis for sub-path $k$ |
| $\theta(N_k,k)$ | Final angle measured from $x$ axis for sub-path $k$ |
| $R_k$ | Radius of the sub-path $k$ |

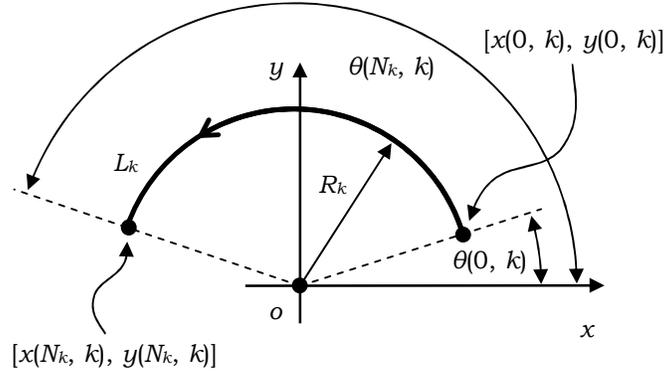

**FIGURE 1**. A semi-circle sub-path $k$ with initial position $x(0,k)$ and $y(0,k)$, length of sub-path $L_k$, initial angle $\theta(0,k)$ and final angle $\theta(N_k,k)$, and radius $R_k$.

*A special case: straight sub-path*

A sub-path with straight form is also a part of semi-circle sub-path but with $R_k$ goes to infinity and $\theta(0,k)=\theta(N_k,k)$. Illustration of this type of path is given in FIGURE 2, with direction of this sub-path is about $\pi/6$ rad.

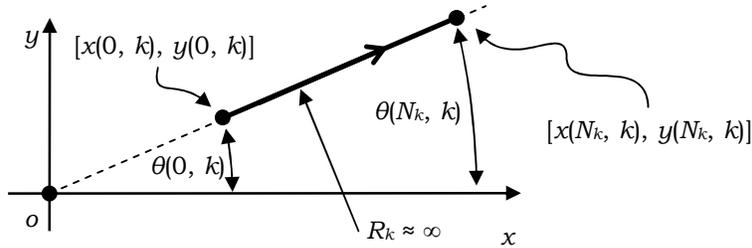

**FIGURE 2**. A straight sub-path $k$ with radius $R_k \approx \infty$.

## Constructing a curved path

A curved path $s$ is constructed using several semi-circle (and also straight) sub-paths, which are discretized in a unit length $\Delta s$. To generate the entire sub-path $k$, a unit length of angle $\Delta \theta_k$ is needed, which is defined as

$$\Delta \theta_k = \frac{\Delta s}{R_k}. \tag{2}$$

EQUATION (2) holds for semi-circle and straight sub-paths. For the last type of sub-path $\Delta \theta_k = 0$. Number of steps is required in constructing the sub-path is given by $N_k$ as defined in EQUATION (1). Position of each segment in a sub-path $k$ is obtained from

$$x(j+1,k) = x(j,k) + \Delta s \cos \theta(j,k), \tag{3}$$

$$y(j+1,k) = y(j,k) + \Delta s \sin \theta(j,k), \tag{4}$$

$$\theta(j+1,k) = \theta(j,k) + \Delta \theta_k, \tag{5}$$

where

$$j = 0 .. N_k - 1. \tag{6}$$

Since last pair of coordinate of previous sub-path will be the first pair of coordinates the next sub-path, then it can written that

$$x(0, k+1) = x(N_k, k), \tag{7}$$

$$y(0, k+1) = y(N_k, k), \tag{8}$$

which conserve the continuity of entire curved path. Then the curve path $s$ can be constructed through EQUATION (3)-(8) in the form of

$$\vec{s} = \sum_{j,k} \Delta \vec{s}(j,k) = \sum_{k=1}^{M-1} \sum_{j=1}^{N_k-1} \left[ \Delta s \cos \theta(j,k) \hat{e}_x + \Delta s \sin \theta(j,k) \hat{e}_y \right], \tag{9}$$

with number of sub-paths is $M$.

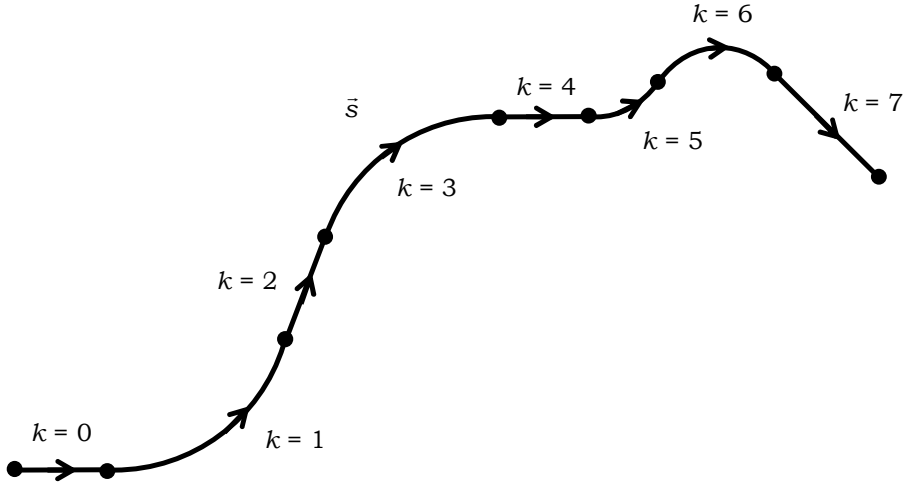

**FIGURE 3**. An example for a curved path $s$ which consists of several sub-paths with $M = 8$.

For the first towo sub-path illustrated in FIGURE 3 the input line can be as follow

```
# Sub-path 0
X0        0
Y0        0
L0        1.5
THETA_A0        0
THETA_B0        0
R0        1E+100

# Sub-path 1
X1        1.5
Y1        0
L1        2.5
THETA_A1        0
THETA_B1        1.046
R1        3
```

which is written in the input file of SCSPG program.

## Considered forces

Some forces are considered in benchmarking the SCSPG program, which are earth gravitational force $\vec{F}_G$, normal force $\vec{F}_N$ from curved path $s$ to a point mass object with mass $m$, and friction force $\vec{F}_S$ in opposite direction of the curved path $s$. Direction of normal force is always perpendicular to the path.

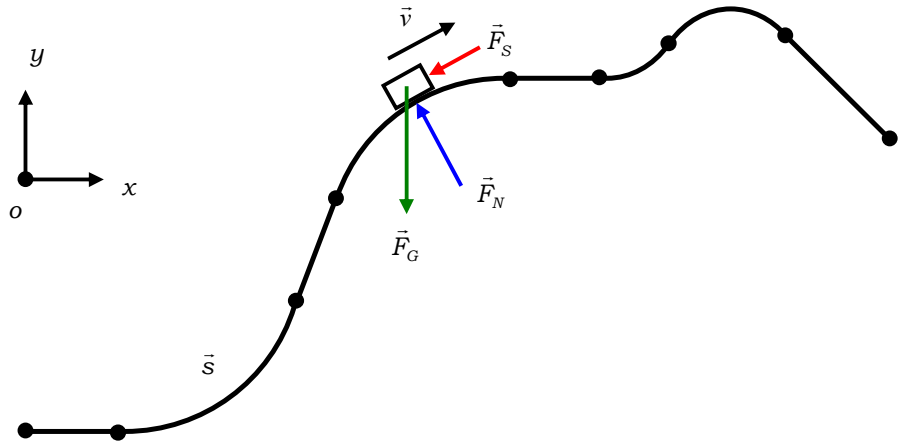

**FIGURE 4**. Considered forces along a curved path $s$.

*Unit vectors*

There are two pairs of unit vectors that will be used in this report. The first pair consists of $\hat{e}_x$ and $\hat{e}_y$, while the second consists of

$$\hat{e}_s = \cos\theta\, \hat{e}_x + \sin\theta\, \hat{e}_y, \tag{10}$$

and

$$\hat{e}_n = -\sin\theta\, \hat{e}_x + \cos\theta\, \hat{e}_y, \tag{11}$$

*Earth gravitational force*

Earth gravitation force can be formulated as

$$\vec{F}_G = -mg\hat{e}_y, \tag{10}$$

with $m$ is mass of the point mass object that travels along the curved path.

*Normal force*

Normal force is defined as

$$\vec{F}_N = F_N \hat{e}_n. \tag{11}$$

Value of $F_N$ is determined from relation in radial direction using concept of sentripetal force

$$F_N = m\left(\frac{v^2}{R} + g\cos\theta\right), \tag{12}$$

where value of $R$ is positive for u-shape and negative fo u-shape semi-circle sub-path. For a straight shape the term $v^2/R$ will be fanished since $R$ goes to infinity.

*Friction force*

Relation between normal force and friction force is given by

$$\vec{F}_S = -\mu_k F_N \frac{\vec{v}}{v}, \tag{13}$$

with

$$\vec{v} = v\hat{e}_s. \tag{14}$$

*Acceleration*

In the direction of curved path $s$ or in direction of velocity the acceleration component will be

$$a_s = \frac{\vec{F}_G \cdot \hat{e}_s - \mu_k F_N}{m} = -\left(g\sin\theta + \frac{\mu_k F_N}{m}\right) \tag{15}$$

and in direction perpendicular to path $s$ or in direction of normal force the acceleration component will be

$$a_n = \frac{F_N - \vec{F}_G \cdot \hat{e}_n}{m} = \frac{F_N}{m} - g\cos\theta = \frac{v^2}{R}. \tag{16}$$

Using right part of EQUATION (16) or EQUATION (12) in EQUATION (15) it can be found that

$$a_s = -g(\sin\theta + \mu_k \cos\theta) - \mu_k \frac{v^2}{R}. \tag{17}$$

EQUATION (17) will be simply reduced to acceleration in incline plane when $R$ goes to infinity.

## Works

Acceleration in EQUATION (17) can be separated into two parts. The first part is due to earth gravitational force and the second is due to friction. Then, the work of friction can be calculated from

$$W_S = \int \vec{F} \cdot d\vec{s} = -\int m\mu_k \left( g\cos\theta + \frac{v^2}{R} \right) \hat{e}_s \cdot \Delta\vec{s}(j,k)\hat{e}_s \cdot d\vec{s} = -\sum_{j,k} m\mu_k \left[ g\cos\theta(j,k) + \frac{v^2(j,k)}{R_k} \right] \hat{e}_s \cdot \Delta\vec{s}(j,k) \quad (17)$$

using EQUATION (9). From EQUATION (3) and (4) it can obtained that

$$v_x(j,k) \approx \frac{x(j+1,k) - x(j,k)}{\Delta t} = \Delta s \cos\theta(j,k), \quad (18)$$

$$v_y(j,k) \approx \frac{y(j+1,k) - y(j,k)}{\Delta t} = \Delta s \sin\theta(j,k), \quad (19)$$

then

$$v^2(j,k) = v_x^2(j,k) + v_y^2(j,k) = \left( \frac{\Delta s}{\Delta t} \right)^2. \quad (20)$$

Result from EQUATION (21) will turn EQUATION (17) to

$$W_S = -\sum_{j,k} m\mu_k \left[ g\cos\theta(j,k) + \frac{(\Delta s/\Delta t)^2}{R_k} \right] \hat{e}_s \cdot \Delta\vec{s}(j,k). \quad (21)$$

Value of $\Delta s/\Delta t$ represents, actually, the velocity of point mass object in every point along the path. At the highest point in a full-circle sub-path, the normal force can be maintained to be equal or larger than zero using this value. Work of friction resulted from SCSPG program will only have physical meaning when

$$\frac{(\Delta s/\Delta t)^2}{R_k} \geq g \quad (22)$$

or

$$\frac{\Delta s}{\Delta t} \geq \sqrt{gR_k}. \quad (23)$$

It means that condition in EQUATION (23) is always required in the input file of SCSPG program.

### Results and discussion

*Incline plane with angle θ*

For a simple incline plane with angle $\theta$ only one sub-path is needed with radius $R$ goes to infinity. Condition of $R$ and single sub-path will reduce EQUATION (21) into

$$W_S = -\sum_j m\mu_k g\cos\theta(j)\hat{e}_s \cdot \Delta\vec{s}(j) = -m\mu_k g\cos\theta \sum_j \Delta s = -m\mu_k gL\cos\theta, \quad (24)$$

which is already well known in common physics texbooks.

*Frictionless roller-coaster (single vertical circular path)*

For single vertical circular path only one sub-path is needed. Then, EQUATION (21) will be reduced into

$$W_S = -\sum_{j,k} m\mu_k \left[ g\cos\theta(j,k) + \frac{(\Delta s/\Delta t)^2}{R_k} \right] \hat{e}_s \cdot \Delta\vec{s}(j,k) = 0, \quad (25)$$

since frictionless path requires $\mu_k = 0$.

*Roller-coaster (single vertical circular path)*

As previously mentioned in EQUATION (25), for this case only one sub-path is needed, which turns EQUATION (21) into

$$W_S = -\sum_j m\mu_k \left[ g\cos\theta(j) + \frac{(\Delta s/\Delta t)^2}{R} \right]\Delta s = -m\mu_k \left[ \frac{L(\Delta s/\Delta t)^2}{R} + g\sum_j \cos\theta(j)\Delta s \right], \quad (26)$$

than can be easily reduced to EQUATION (24) and (25) for each previous condition, $\theta(j) = \theta$ and $R$ goes to infinity, and $\mu_k = 0$, respectively.

*Common curved path*

A curved path can be composed into several straight and semi-circle sub-paths as already given in EQUATION (24) and (26), where an example is given in TABLE 1 with the input data for the SCSPG program.

**TABLE 2**. Input example of SCSPG program and its result.

| Input | Result |
|---|---|
| # Path discretization<br>DS    1E-3<br><br># Time discretization<br>DT    1E-3<br><br># Number of sub-paths<br>N     2<br><br># Sub-path 0<br>X0     0<br>Y0     0<br>L0     3<br>THETA0 0.524<br>R0     1E+100<br><br># Sub-path 1<br>X1     2.598<br>Y1     1.5<br>L1     6.28<br>THETA1 0.524<br>R1     2E+0 | (plot of y (m) vs x (m) showing the curved path) |

It can be seen in TABLE 2 that the curved path consists of two sub-paths: straight path and semi-circle path. The first sub-path starts from (0, 0) with $\theta_0 = 0.524$ rad, $L_0 = 3$, and $R_0 \approx \infty$, while the second starts from (2.598, 1.5) with $\theta_1 = 0.524$ rad, $L_1 = 6.28$, and $R_1 = 2$.

*Calculated work of friction*

For a straight sub-path value of $\Delta s/\Delta t$ does not require to obey EQUATION (23) but it does require to obey it for a semi-circular path. For the first case with $\theta = 0.524$, $L = 3$, $R \approx \infty$, $m = 1$, $\mu_k = 0.2$, and $g = 10$, it has been found that the work of friction calculated from SCSPG program is independent of $\Delta s/\Delta t$. Calculated value is -3.4633 while the theoretical prediction is -3.4632. It is only about 0.003 % difference. Unfortunately it is quite different for the second case, where work of fricton is given in FIGURE 5. The path is similar to the second path as given in TABLE 2.

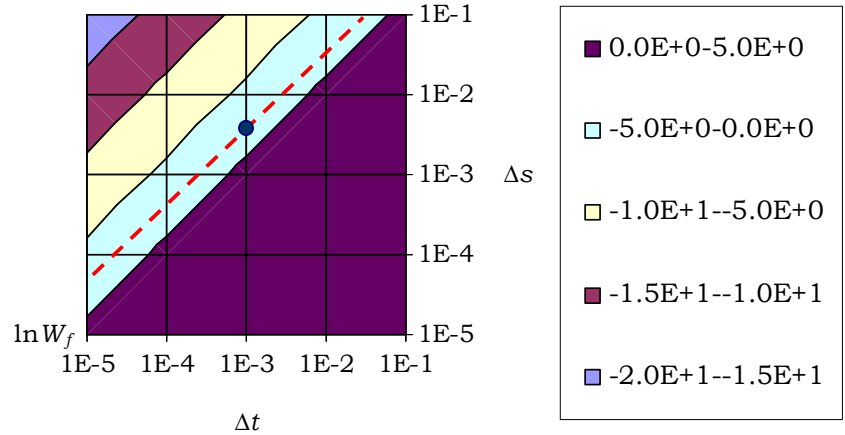

**FIGURE 5**. Work of friction $W_f$ as function of $\Delta s$ and $\Delta t$.

Since work of friction must be less than zero and $\Delta s/\Delta t$ > 4.472 from EQUATION (23), then it can be found from FIGURE 5 the cyan area (-5 until 0) already represent work of friction which has physical meaning. A blue circle dot on the dashed red line gives $W_f < -7.389$.

## Conclusion

SCSPG program is reported in this work, which is able to draw a curved path consisted of several sub-path of semi-circle or straight type and is able to calculate the work of friction along the curved path with minimum $\Delta s/\Delta t$ is required to produce result with physical meaning.

## Acknowledgements

Author S.V. would like to thank ITB Alumni Association Research Grant (HR-IA-ITB) and ITB Innovation and Research Division Research Grant (RIK-ITB) in year 2011-2012 for supporting computation part of this work.